\documentclass[letterpaper,twocolumn,10pt]{article}
\usepackage{usenix,epsfig,endnotes}
\usepackage{times}
\usepackage{balance}
\usepackage{amsfonts}
\usepackage{authblk}
\usepackage{epsfig}  
\usepackage{amssymb}
\usepackage[normalem]{ulem}
\usepackage{amscd}
\usepackage{amsmath} 
\usepackage{microtype}
\usepackage{array}
\usepackage{chngpage}
\usepackage{graphics}
\usepackage{epsfig}
\usepackage{multirow}
\usepackage{graphicx}
\usepackage{caption}
\usepackage{listings}
\usepackage{subfigure}
\usepackage{adjustbox}
\usepackage[hyphens]{url}
\usepackage{tabularx}
\usepackage{epstopdf}
\usepackage{threeparttable}
\usepackage{ulem}
\usepackage{lipsum}
\usepackage{float}
\usepackage{pifont}
\usepackage{bm}
\usepackage{booktabs}
\usepackage{color,xcolor}
\usepackage{booktabs} 
\usepackage{verbatim}
\usepackage{amssymb}
\usepackage{pifont}
\usepackage{array,multirow}
\usepackage{makecell}
\usepackage{tikz}
\usepackage{CJK}
\usepackage{appendix}  
\usepackage{float}

\newcommand{\ignore}[1]{}
\newcommand{\revised}[1]{}

\newcommand\Shengzhi[1]{{\textcolor{red}{#1}}}

\renewcommand{\thefootnote}{\fnsymbol{footnote}}
\title{\Large \bf CommanderSong: A Systematic Approach for  Practical Adversarial Voice Recognition}

\author[1,2]{\normalsize Xuejing Yuan}
\author[3]{Yuxuan Chen}
\author[1,2]{Yue Zhao}
\author[4]{Yunhui Long}
\author[1,2]{Xiaokang Liu}
\author[1,2]{Kai Chen\thanks{Corresponding author: chenkai@iie.ac.cn}  }
\author[3,5]{Shengzhi Zhang} 
\author[ ]{\\Heqing Huang}
\author[6]{XiaoFeng Wang}
\author[4]{Carl A. Gunter}
\affil[1]{\normalsize{SKLOIS, Institute of Information Engineering, Chinese Academy of Sciences, China}}
\affil[2]{School of Cyber Security, University of Chinese Academy of Sciences, China}
\affil[3]{Department of Computer Science, Florida Institute of Technology, USA}
\affil[4]{Department of Computer Science, University of Illinois at Urbana-Champaign, USA}
\affil[5]{Department of Computer Science, Metropolitan College, Boston University, USA}
\affil[6]{School of Informatics and Computing, Indiana University Bloomington, USA}

\begin{document}
\maketitle 
\renewcommand{\thefootnote}{\arabic{footnote}}
\pagestyle{empty}

\subsection*{Abstract}
The popularity of automatic speech recognition (ASR) systems, like Google Assistant, Cortana, brings in security concerns, as demonstrated by recent attacks.  The impacts of such threats, however, are less clear, since they are either less stealthy (producing noise-like voice commands) or requiring the physical presence of an attack device (using ultrasound speakers or transducers). In this paper, we demonstrate that not only are more practical and surreptitious attacks feasible but they can even be \textit{automatically} constructed. Specifically, we find that the voice commands can be stealthily embedded into songs, which, when played, can effectively control the target system through ASR without being noticed. For this purpose, we developed novel techniques that address a key technical challenge: integrating the commands into a song in a way that can be effectively recognized by ASR through the air, in the presence of background noise, while not being detected by a human listener. Our research shows that this can be done automatically against real world ASR applications\footnote{Demos of attacks are uploaded on the website
(https://sites.google.com/view/commandersong/)}. We also demonstrate that such \textit{CommanderSongs} can be spread through Internet (e.g., YouTube) and radio, potentially affecting millions of ASR users. Finally we present mitigation techniques that defend existing ASR systems against such threat.

\section{Introduction}

\label{sec:intro}

Intelligent voice control (IVC) has been widely used in human-computer interaction, such as Amazon Alexa~\cite{alexa}, Google Assistant~\cite{googleassistant}, Apple Siri~\cite{apple}, Microsoft Cortana~\cite{microsoft} and iFLYTEK~\cite{iFLYTEK}. Running the state-of-the-art ASR techniques, these systems can effectively interpret natural voice commands and execute the corresponding operations such as unlocking the doors of home or cars, making online purchase, sending messages, and etc. This has been made possible by recent progress in machine learning, deep learning~\cite{hinton2012deep} in particular, which vastly improves the accuracy of speech recognition. In the meantime, these deep learning techniques are known to be vulnerable to adversarial perturbations~\cite{kurakin2016adversarial,brown2017adversarial,evtimov2017robust,dalvi2004adversarial,biggio2013evasion,szegedy2013intriguing,goodfellow2014explaining,papernot2016transferability}. Hence, it becomes imperative to understand the security implications of the ASR systems in the presence of such attacks. 

\vspace{3pt}\noindent\textbf{Threats to ASR} Prior research shows that carefully-crafted perturbations, even a small amount, could cause a machine learning classifier to misbehave in an unexpected way. Although such adversarial learning has been extensively studied in image recognition, little has been done in speech recognition, potentially due to the new challenge in this domain: unlike adversarial images, which include the perturbations of less noticeable background pixels, changes to voice commands often introduce noise that a modern ASR system is designed to filter out and therefore cannot be easily misled. 

Indeed, a recent attack on ASR utilizes noise-like hidden voice command~\cite{carlini2016hidden}, but the white box attack is based on a traditional speech recognition system that uses a Gaussian Mixture Model (GMM), not the DNN behind today's ASR systems. Another attack transmits inaudible commands through ultrasonic sound~\cite{zhang2017dolphinattack}, but it exploits microphone hardware vulnerabilities instead of the weaknesses of the DNN. Moreover, an attack device, e.g., an ultrasonic transducer or speaker, needs to be placed close to the target ASR system. So far little success has been reported in generating ``adversarial sound'' that practically fools deep learning technique but remains inconspicuous to human ears, and meanwhile allows it to be played from the remote (e.g., through YouTube) to attack a large number of ASR systems. 

To find \textit{practical} adversarial sound, a few technical challenges need to be addressed: (C1) the adversarial audio sample is expected to be effective in a complicated, real-world audible environment, in the presence of electronic noise from speaker and other noises; (C2) it should be stealthy, unnoticeable to ordinary users; (C3) impactful adversarial sound should be remotely deliverable and can be played by popular devices from online sources, which can affect a large number of IVC devices. All these challenges have been found in our research to be completely addressable, indicating that the threat of audio adversarial learning is indeed realistic.

\vspace{3pt}\noindent\textbf{CommanderSong.} More specifically, in this paper, we report a practical and systematic adversarial attack on real world speech recognition systems. Our attack can \textit{automatically} embed a set of commands into a (randomly selected) song, to spread to a large amount of audience (addressing C3). This revised song, which we call \textit{CommanderSong}, can sound completely normal to ordinary users, but will be interpreted as commands by ASR, leading to the attacks on real-world IVC devices.
To build such an attack, we leverage an open source ASR system Kaldi~\cite{kaldi}, which includes acoustic model and language model. By carefully synthesizing the outputs of the acoustic model from both the song and the given voice command, we are able to generate the adversarial audio with minimum perturbations through gradient descent, so that the CommanderSong can be less noticeable to human users (addressing C2, named WTA attack). To make such adversarial samples practical, our approach has been designed to capture the electronic noise produced by different speakers, and integrate a generic noise model into the algorithm for seeking adversarial samples (addressing C1, called WAA attack).

In our experiment, we generated over 200 CommanderSongs that contain different commands, and attacked Kaldi with an 100\% success rate in a WTA attack and a 96\% success rate in a WAA attack. Our evaluation further demonstrates that such a CommanderSong can be used to perform a \textit{black box} attack on a mainstream ASR system iFLYTEK\footnote{We have reported this to iFLYTEK, and are waiting for their responses.}~\cite{iFLYTEK} (neither source code nor model is available). iFLYTEK has been used as the voice input method by many popular commercial apps, including WeChat (a social app with 963 million users), Sina Weibo (another social app with 530 million users), JD (an online shopping app with 270 million users), etc. To demonstrate the impact of our attack, we show that CommanderSong can be spread through YouTube, which might impact millions of users. To understand the human perception of the attack, we conducted a user study\footnote{The study is approved by the IRB.} on Amazon Mechanical Turk~\cite{AmazonTurk}. Among over 200 participants, none of them identified the commands inside our CommanderSongs. 
We further developed the defense solutions against this attack and demonstrated their effectiveness.

\vspace {5pt}\noindent\textbf{Contributions}. The contributions of this paper are summarized as follows:

\vspace {2pt}\noindent$\bullet$\space\textit{Practical adversarial attack against ASR systems}. We designed and implemented the first practical adversarial attacks against ASR systems. \ignore{To the best of our knowledge, this is the first systematical approach to generate such practical attacks against DNN-based ASR systems.} Our attack is demonstrated to be \textit{robust}, working across air in the presence of environmental interferences, \textit{transferable}, effective on a black box commercial ASR system (i.e., iFLYTEK) and \textit{remotely deliverable}, potentially impacting millions of users.

\vspace {2pt}\noindent$\bullet$\space\textit{Defense against CommanderSong.} We design two approaches (audio turbulence and audio squeezing) to defend against the attack, which proves to be effective by our preliminary experiments.

\vspace {5pt}\noindent\textbf{Roadmap}. The rest of the paper is organized as follows: Section~\ref{sec:background} gives the background information of our study. Section~\ref{sec:overview} provides motivation and overviews our approach. In Section~\ref{sec:attack}, we elaborate the design and implementation of CommanderSong. In Section~\ref{sec:Evaluation}, we present the experimental results, with emphasis on the difference between machine and human comprehension. Section~\ref{sec:understanding} investigates deeper understanding on CommanderSongs. Section~\ref{sec:Defense} shows the defense of the CommanderSong attack. Section~\ref{sec:relate} compares our work with prior studies and Section~\ref{sec:Conclusion} concludes the paper.

\section{Background}
\label{sec:background}
In this section, we overview existing speech recognition system,  
and discuss the recent advance on the attacks against both image and speech recognition systems. 

\begin{figure*}[t]
\centering
\includegraphics[width=0.8\textwidth]{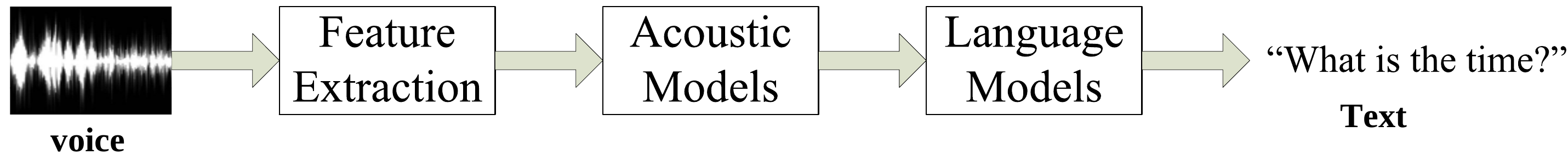}
\caption{Architecture of Automatic Speech Recognition System.} 
\label{speech}
\end{figure*}

\subsection{Speech Recognition}
Automatic speech recognition is a technique that allows machines to recognize/understand the semantics of human voice. Besides the commercial products like Amazon Alexa, Google Assistant, Apple Siri, iFLYTEK, etc., there are also open-source platforms such as Kaldi toolkit~\cite{kaldi}, Carnegie Mellon University's Sphinx toolkit~\cite{cmu}, HTK toolkit~\cite{htk}, etc. Figure~\ref{speech} presents an overview of a typical speech recognition system, with two major components: feature extraction and decoding based on pre-trained models (e.g., acoustic models and language models).

After the raw audio is amplified and filtered, acoustic features need to be extracted from the preprocessed audio signal. The features contained in the signal change significantly over time, so short-time analysis is used to evaluate them periodically. Common acoustic feature extraction algorithms include Mel-Frequency Cepstral Coefficients (MFCC)~\cite{muda2010voice}, Linear Predictive Coefficient (LPC)~\cite{itakura1975line}, Perceptual Linear Predictive (PLP)~\cite{hermansky1990perceptual}, etc. Among them, MFCC is the most frequently used one in both open source toolkit and commercial products~\cite{o2008automatic}. GMM can be used to analyze the property of the acoustic features. The extracted acoustic features are matched against pre-trained acoustic models to obtain the likelihood probability of phonemes. Hidden Markov Models (HMM) are commonly used for statistical speech recognition. As GMM is limited to describe a non-linear manifold of the data, Deep Neural Network-Hidden Markov Model (DNN-HMM) has been widely used for speech recognition in academic and industry community since 2012~\cite{DNN-HMM}. 

Recently, end-to-end deep learning becomes used in speech recognition systems. It applies a large scale dataset and uses CTC (Connectionist Temporal Classification) loss function to directly obtain the characters rather than phoneme sequence. CTC locates the alignment of text transcripts with input speech using an all-neural, sequence-to-sequence neural network. Traditional speech recognition systems involve many engineered processing stages, while CTC can supersede these processing stages via deep learning~\cite{DeepSpeech}. The architecture of end-to-end ASR systems always includes an encoder network corresponding to the acoustic model and a decoder network corresponding to the language model~\cite{End-to-end}. DeepSpeech~\cite{DeepSpeech} and Wav2Letter~\cite{Wav2Letter} are popular open source end-to-end speech recognition systems. 

\subsection{Existing Attacks against Image and Speech Recognition Systems}
Nowadays people are enjoying the convenience of integrating image and speech as new input methods into mobile devices. Hence, the accuracy and dependability of image and speech recognition pose critical impact on the security of such devices. Intuitively, the adversaries can compromise the integrity of the training data if they have either physical or remote access to it. By either revising existing data or inserting extra data in the training dataset, the adversaries can certainly tamper the dependability of the trained models~\cite{BEBP}. 

When adversaries do not have access to the training data, attacks are still possible. Recent research has been done to deceive image recognition systems into making wrong decision by slightly revising the input data. The fundamental idea is to revise an image slightly to make it ``look'' different from the views of human being and machines. Depending on whether the adversary knows the algorithms and parameters used in the recognition systems, there exist white box and black box attacks. Note that the adversary always needs to be able to interact with the target system to observe corresponding output for any input, in both white and black box attacks. Early researches~\cite{tronci2011fusion,schuckers2002spoofing,bao2009liveness} focus on the revision and generation of the digital image file, which is directly fed into the image recognition systems. The state-of-the-art researches~\cite{kurakin2016adversarial,brown2017adversarial,evtimov2017robust} advance in terms of practicality by printing the adversarial image and presenting it to a device with image recognition functionality. 

However, the success of the attack against image recognition systems has not been ported to the speech recognition systems until very recently, due to the complexity of the latter. The speech, a time-domain continuous signal, contains much more features compared to the static images. Hidden voice command~\cite{carlini2016hidden} launched both black box (i.e., inverse MFCC) and white box (i.e., gradient decent) attacks against speech recognition systems, and generated obfuscated commands to ASR systems. Though seminal in attacking speech recognition systems, it is also limited to make practical attacks. For instance, a large amount of human effort is involved as feedback for the black box approach, and the white box approach is based on GMM-based acoustic models, which have been replaced by DNN-based ones in most modern speech recognition systems. The recent work DolphinAttack~\cite{zhang2017dolphinattack} proposed a completely inaudible voice attack by modulating commands on ultrasound carriers and leveraging microphone vulnerabilities (i.e., the nonlinearity of the microphones). As noted by the authors, such attack can be eliminated by an enhanced microphone that can suppress acoustic signals on ultrasound carrier, like iPhone 6 Plus. 

\section{Overview}
\label{sec:overview}

In this section, we present the motivation of our work, and overview the proposed approach to generate the practical adversarial attack. 

\subsection{Motivation}
Recently, adversarial attacks on image classification have been extensively studied~\cite{brown2017adversarial,evtimov2017robust}. Results show that even the state-of-the-art DNN-based classifier can be fooled by small perturbations added to the original image~\cite{kurakin2016adversarial}, producing erroneous classification results. However, the impact of adversarial attacks on the most advanced speech recognition systems, such as those integrating DNN models, has never been systematically studied. Hence, in this paper, we investigated DNN-based speech recognition systems, and explored adversarial attacks against them. Researches show that commands can be transmitted to IVC devices through inaudible ultrasonic sound~\cite{zhang2017dolphinattack} and noises~\cite{carlini2016hidden}. Even though the existing works against ASR systems are seminal, they are limited in some aspects. Specifically, ultrasonic sound can be defeated by using a low-pass filter (LPF) or analyzing the signal frequency range, and noises are easy to be noticed by users. 

Therefore, the research in this paper is motivated by the following questions: (Q1) Is it possible to build the practical adversarial attack against ASR systems, given the facts that the most ASR systems are becoming more intelligent (e.g., by integrating DNN models) and that the generated adversarial samples should work in the very complicated physical environment, e.g., electronic noise from speaker, background noise, etc.? (Q2) Is it feasible to generate the adversarial samples (including the target commands) that are difficult, or even impossible, to be noticed by ordinary users, so the control over the ASR systems can happen in a ``hidden'' fashion? (Q3) If such adversarial audio samples can be produced, is it possible to impact a large amount of victims in an automated way, rather than solely relying on attackers to play the adversarial audio and affecting victims nearby? Below, we will detail how our attack is designed to address the above questions.

\subsection{The Philosophy of Designing Our Attack}
To address Q3, our idea is to choose songs as the ``carrier'' of the voice commands recognizable by ASR systems. The reason of choosing such ``carrier'' is at least two-fold. On one hand, enjoying songs is always a preferred way for people to relax, e.g., listening to the music station, streaming music from online libraries, or just browsing YouTube for favorite programs. Moreover, such entertainment is not restricted by using radio, CD player, or desktop computer any more. A mobile device, e.g., Android phone or Apple iPhone, allows people to enjoy songs everywhere. Hence, choosing the song as the ``carrier'' of the voice command automatically helps impact millions of people. On the other hand, ``hiding'' the desired command in the song also makes the command much more difficult to be noticed by victims, as long as Q2 can be reasonably addressed. Note that we do not rely on the lyrics in the song to help integrate the desired command. Instead, we intend to avoid the songs with the lyrics similar to our desired command. For instance, if the desired command is ``open the door'', choosing a song with the lyrics of ``open the door'' will easily catch the victims' attention. 
Hence, we decide to use random songs as the ``carrier'' regardless of the desired commands.

Actually choosing the songs as the ``carrier'' of desired commands makes Q2 even more challenging. Our basic idea is when generating the adversarial samples, we revise the original song leveraging the pure voice audio of the desired command as a reference. In particular, we find the revision of the original song to generate the adversarial samples is always a trade off between preserving the fidelity of the original song and recognizing the desired commands from the generated sample by ASR systems. To better obfuscate the desired commands in the song, in this paper we emphasize the former than the latter. In other words, we designed our revision algorithm to maximally preserve the fidelity of the original song, at the expense of losing a bit success rate of recognition of the desired commands. However, such expense can be compensated by integrating the same desired command multiple times into one song (the command of ``open the door'' may only last for 2 seconds.), and the successful recognition of one suffices to impact the victims.

Technically, in order to address Q2, we need to investigate the details of an ASR system. As shown in Figure \ref{speech}, an ASR system is usually composed of two pre-trained models: an acoustic model describing the relationship between audio signals and phonetic units, and a language model representing statistical distributions over sequences of words. In particular, given a piece of pure voice audio of the desired command and a ``carrier'' song, we can feed them into an ASR system separately, and intercept the intermediate results. By investigating the output from the acoustic model when processing the audio of the desired command, and the details of the language model, we can conclude the ``information'' in the output that is necessary for the language model to produce the correct text of the desired command. When we design our approach, we want to ensure such ``information'' is only a small subset (hopefully the minimum subset) of the output from the acoustic model. Then, we carefully craft the output from the acoustic model when processing the original song, to make it ``include'' such ``information'' as well. Finally, we inverse the acoustic model and the feature extraction together, to directly produce the adversarial sample based on the crafted output (with the ``information'' necessary for the language model to produce the correct text of the desired command).

Theoretically, the adversarial samples generated above can be recognized by the ASR systems as the desired command if directly fed as input to such systems. Since such input usually is in the form of a wave file (in ``WAV" format) and the ASR systems need to expose APIs to accept the input, we define such attack as the WAV-To-API (WTA) attack. However, to implement a practical attack as in Q1, the adversarial sample should be played by a speaker to interact with IVC devices over the air. In this paper, we define such practical attack as WAV-Air-API (WAA) attack. The challenge of the WAA attack is when playing the adversarial samples by a speaker, the electronic noise produced by the loudspeakers and the background noise in the open air have significant impact on the recognition of the desired commands from the adversarial samples. To address this challenge, we improve our approach by integrating a generic noise model to the above algorithm with the details in Section \ref{WAA}.

\section{Attack Approach}
\label{sec:attack}
We implement our attack by addressing two technical challenges: (1) minimizing the perturbations to the song, so the distortion between the original song and the generated adversarial sample can be as unnoticeable as possible, and (2) making the attack practical, which means CommanderSong should be played over the air to compromise IVC devices. To address the first challenge, we proposed \textit{pdf-id sequence matching} to incur minimum revision at the output of the acoustic model, and use gradient descent to generate the corresponding adversarial samples as in Section \ref{WTA}. The second challenge is addressed by introducing a generic noise model to simulate both the electronic noise and background noise as in Section \ref{WAA}. Below we elaborate the details.

\begin{figure}[t]
\centering
\includegraphics[width=0.45\textwidth]{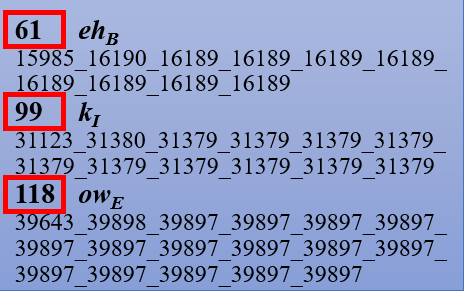}
\caption{Result of decoding ``Echo". }
\label{decode lat}
\end{figure}

\begin{table}[t]
\centering
\caption{Relationship between transition-id and pdf-id.}
\label{Transition-id with pdf-id}

\begin{tabular}{m{1.2cm}<{\centering}|m{0.9cm}<{\centering}|m{0.6cm}<{\centering}|m{1.5cm}<{\centering}|m{1.5cm}<{\centering}} 
\Xhline{3\arrayrulewidth}
\textbf{Phoneme} & \textbf{HMM-state} & \textbf{Pdf-id} & \textbf{Transition-id} & \textbf{Transition} \\ 
\Xhline{3\arrayrulewidth}
\multirow{2}{*}{$eh_B$}& \multirow{2}{*}{0} & \multirow{2}{*}{6383} & \text{15985} & \text{0$\rightarrow$1} \\ \cline{4-5} & & & \text{15986}& \text{0$\rightarrow$2} \\ \hline
\multirow{2}{*}{$eh_B$} & \multirow{2}{*}{1} & \multirow{2}{*}{5760} & \text{16189} &\text{self-loop} \\ \cline{4-5} & & & \text{16190}& {1$\rightarrow$2} \\ \hline
\multirow{2}{*}{$k_I$} & \multirow{2}{*}{0} & \multirow{2}{*}{6673} &\text{31223}&\text{0$\rightarrow$1}\\ \cline{4-5} 
& & & \text{31224}& \text{0$\rightarrow$2}  \\ \hline
\multirow{2}{*}{$k_I$} &\multirow{2}{*}{1} &\multirow{2}{*}{3787} & \text{31379}& \text{self-loop}\\  \cline{4-5} & & & \text{31380} & \text{1$\rightarrow$2} \\ \hline
\multirow{2}{*}{$ow_E$} &\multirow{2}{*}{0} &\multirow{2}{*}{5316} & \text{39643}& \text{0$\rightarrow$1}\\ \cline{4-5} & & & \text{9644}& \text{0$\rightarrow$2}  \\ \hline
\multirow{2}{*}{$ow_E$} &\multirow{2}{*}{1} &\multirow{2}{*}{8335} & \text{39897}& \text{self-loop}\\ \cline{4-5} & & & \text{39898} & \text{1$\rightarrow$2}  \\ 
\Xhline{3\arrayrulewidth}
\end{tabular}
\end{table}

\subsection{Kaldi Platform}
\label{preparation}
We choose the open source speech recognition toolkit Kaldi~\cite{kaldi}, due to its popularity in research community. Its source code on github obtains 3,748 stars and 1,822 forks~\cite{Aspiremodel}. Furthermore, the corpus trained by Kaldi on ``Fisher" is also used by IBM~\cite{Fisher_IBM} and Microsoft~\cite{Fisher_Microsoft}. 

In order to use Kaldi to decode audio, we need a trained model to begin with. There are some models on Kaldi website that can be used for research. We took advantage of the ``ASpIRE Chain Model" (referred as ``ASpIRE model'' in short), which was one of the latest released decoding models when we began our study\footnote{There are three decoding models on Kaldi platform currently. ASpIRE Chain Model we used in this paper was released on October 15th, 2016, while SRE16 Xvector Model was released on October 4th, 2017, which was not available when we began our study. The CVTE Mandarin Model, released on June 21st 2017 was trained in Chinese~\cite{kaldi}.}. After manually analyzing the source code of Kaldi (about 301,636 lines of shell scripts and 238,107 C++ SLOC), we completely explored how Kaldi processes audio and decodes it to texts. Firstly, it extracts acoustic features like MFCC or PLP from the raw audio. Then based on the trained probability density function (p.d.f.) of the acoustic model, those features are taken as input to DNN to compute the posterior probability matrix. The p.d.f. is indexed by the pdf identifier (pdf-id), which exactly indicates the column of the output matrix of DNN. 

Phoneme is the smallest unit composing a word. There are three states (each is denoted as an  HMM state) of sound production for each phoneme, and a series of transitions among those states can identify a phoneme. A transition identifier (transition-id) is used to uniquely identify the HMM state transition. Therefore, a sequence of transition-ids can identify a phoneme, so we name such a sequence as \textit{phoneme identifier} in this paper. Note that the transition-id is also mapped to pdf-id. Actually, during the procedure of Kaldi decoding, the phoneme identifiers can be obtained. By referring to the pre-obtained mapping between transition-id and pdf-id, any phoneme identifier can also be expressed as a specific sequence of pdf-ids. Such a specific sequence of pdf-ids actually is a segment from the posterior probability matrix computed from DNN. This implies that to make Kaldi decode any specific phoneme, we need to have DNN compute a posterior probability matrix containing the corresponding sequence of pdf-ids.

To illustrate the above findings, we use Kaldi to process a piece of audio with several known words, and obtain the intermediate results, including the posterior probability matrix computed by DNN, the transition-ids sequence, the phonemes, and the decoded words. Figure 2 demonstrates the decoded result of \textit{Echo}, which contains three phonemes. The red boxes highlight the id representing the corresponding phoneme, and each phoneme is identified by a sequence of transition-ids, or the \textit{phoneme identifiers}. Table~\ref{Transition-id with pdf-id} is a segment from the the relationship among the phoneme, pdf-id, transition-id, etc. By referring to Table~\ref{Transition-id with pdf-id}, we can obtain the pdf-ids sequence corresponding to the decoded transition-ids sequence\footnote{For instance, the pdf-ids sequence for $eh_B$ should be \textit{6383, 5760, 5760, 5760, 5760, 5760, 5760, 5760, 5760, 5760}.}. Hence, for any posterior probability matrix demonstrating such a pdf-ids sequence should be decoded by Kaldi as $eh_B$.

\begin{figure*}[t]
\centering
\includegraphics[width=0.8\textwidth]{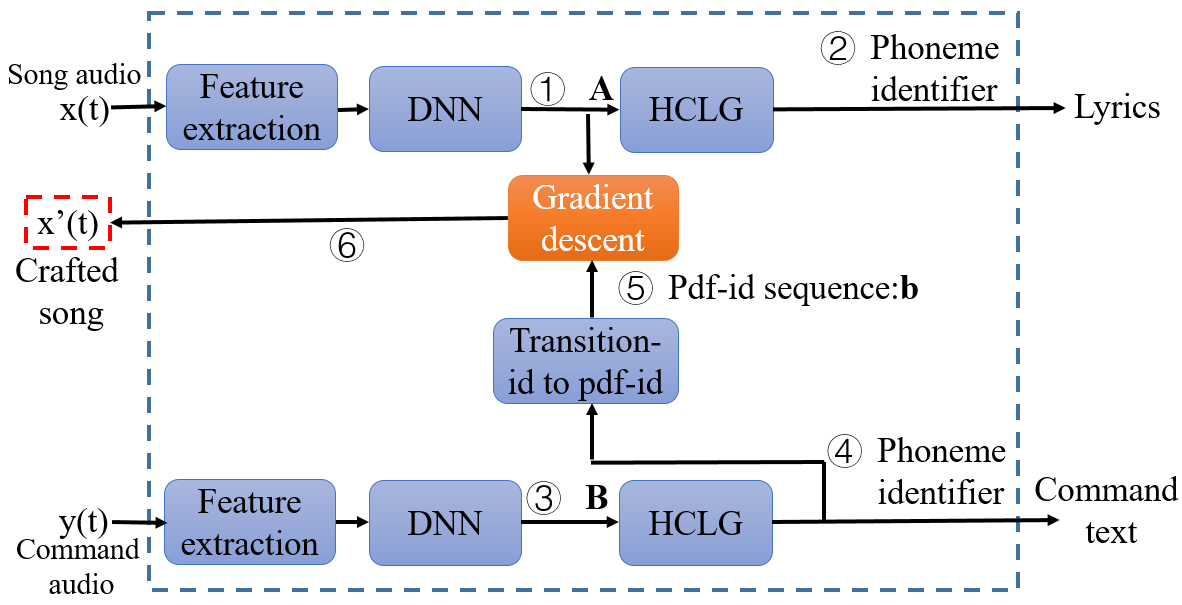}
\caption{Steps of attack. \label{overflow}}
\end{figure*}  

\subsection{Gradient Descent to Craft Audio}
\label{WTA}
Figure~\ref{overflow} demonstrates the details of our attack approach. Given the original song $x(t)$ and the pure voice audio of the desired command $y(t)$, we use Kaldi to decode them separately. By analyzing the decoding procedures, we can get the output of DNN matrix $\textbf{A}$ of the original song (Step \textcircled{1} in Figure~\ref{overflow}) and the phoneme identifiers of the desired command audio (Step \textcircled{4} in Figure~\ref{overflow}). 

The DNN's output $A$ is a matrix containing the probability of each pdf-id at each frame. Suppose there are $n$ frames and $k$ pdf-ids, let $a_{i,j}$ ($1 \leq i\leq n, 1 \leq j \leq k$) be the element at the $i$th row and $j$th column in $\textbf{A}$. Then $a_{i,j}$ represents the probability of the $jth$ pdf-id at frame $i$. For each frame, we calculate the most likely pdf-id as the one with the highest probability in that frame. That is,
\begin{equation*}
	m_i = \arg\max_{j}a_{i,j}.
\end{equation*}
Let $\mathbf{m} = (m_1, m_2, \dots, m_n)$. $\mathbf{m}$ represents a sequence of most likely pdf-ids of the original song audio $x(t)$. For simplification, we use $g$ to represent the function that takes the original audio as input and outputs a sequence of most likely pdf-ids based on DNN's predictions. That is,
\begin{equation*}
	g(x(t)) = \mathbf{m}.
\end{equation*}

As shown in Step~\textcircled{5} in Figure~\ref{overflow}, we can extract a sequence of pdf-id of the command $\mathbf{b} = (b_1, b_2, \dots, b_n)$, where $b_i$ ($1\leq i \leq n$) represents the highest probability pdf-id of the command at frame $i$. To have the original song decoded as the desired command, we need to identify the minimum modification $\delta(t)$ on $x(t)$ so that $\mathbf{m}$ is same or close to $\mathbf{b}$. Specifically, we minimize the $L1$ distance between $\mathbf{m}$ and $\mathbf{b}$. As $\mathbf{m}$ and $\mathbf{b}$ are related with the pdf-id sequence, we define this method as \textit{pdf-id sequence matching} algorithm.

Based on these observations we construct the following objective function:
\begin{equation} \label{eq:obj}
	\arg\min_{\delta(t)} \quad \lVert g\left(x(t) + \delta(t)\right) - \mathbf{b} \rVert_1.
\end{equation}
To ensure that the modified audio does not deviate too much from the original one, we optimize the objective function Eq (\ref{eq:obj}) under the constraint of $\left|\delta(t)\right|\leq l$. 

Finally, we use gradient descent~\cite{papernot2016cleverhans}, an iterative optimization algorithm to find the local minimum of a function, to solve the objective function. Given an initial point, gradient descent follows the direction which reduces the value of the function most quickly. By repeating this process until the value starts to remain stable, the algorithm is able to find a local minimum value. In particular, based on our objective function, we revise the song $x(t)$ into $x'(t) = x(t) + \delta(t)$ with the aim of making most likely pdf-ids $g\left(x'(t)\right)$ equal or close to $\mathbf{b}$. Therefore, the crafted audio $x'(t)$ can be decoded as the desired command. 

To further preserve the fidelity of the original song, one method is to minimize the time duration of the revision. Typically, once the pure command voice audio is generated by a text-to-speech engine, all the phonemes are determined, so as to the phoneme identifiers and $\mathbf{b}$. However, the speed of the speech also determines the number of frames and the number of transition-ids in a phoneme identifier. Intuitively, slow speech always produces repeated frames or transition-ids in a phoneme. Typically people need six or more frames to realize a phoneme, but most speech recognition systems only need three to four frames to interpret a phoneme. Hence, to introduce the minimal revision to the original song, we can analyze $\mathbf{b}$, reduce the number of repeated frames in each phoneme, and obtain a shorter $\mathbf{b}'=(b_1, b_2, \dots, b_q)$, where $q<n$.

\subsection{Practical Attack over the Air}
\label{WAA}
By feeding the generated adversarial sample directly into Kaldi, the desired command can be decoded correctly. However, playing the sample through a speaker to physically attack an IVC device typically cannot work. This is mainly due to the noises introduced by the speaker and environment, as well as the distortion caused by the receiver of the IVC device. In this paper, we do not consider the invariance of background noise in different environments, e.g., grocery, restaurant, office, etc., due to the following reasons: (1) In a quite noisy environment like restaurant or grocery, even the original voice command $y(t)$ may not be correctly recognized by IVC devices; (2) Modeling any slightly variant background noise itself is still an open research problem; (3) Based on our observation, in a normal environment like home, office, lobby, the major impacts on the physical attack are the electronic noise from the speaker and the distortion from the receiver of the IVC devices, rather than the background noise.

Hence, our idea is to build a noise model, considering the speaker noise, the receiver distortion, as well as the generic background noise, and integrate it in the approach in Section \ref{WTA}. Specifically, we carefully picked up several songs and played them through our speaker in a very quiet room. By comparing the recorded audio (captured by our receiver) with the original one, we can capture the noises. Note that playing ``silent'' audio does not work since the electronic noise from speakers may depend on the sound at different frequencies. Therefore, we intend to choose the songs that cover more frequencies. Regarding the comparison between two pieces of audio, we have to first manually align them and then compute the difference.
\ignore{subtract each value in the first WAV file from that in the second one. The result of subtraction is the captured noise.} 
We redesign the objective function as shown in Eq~\eqref{eq:modify}.
\begin{equation} \label{eq:modify}
	\arg\min_{\mu(t)} \quad \lVert g\left(x(t) + \mu(t)+n(t)\right) - \mathbf{b} \rVert _1,
\end{equation}
where $\mu(t)$ is the perturbation that we add to the original song, and $n(t)$ is the noise samples that we captured. In this way, we can get the adversarial audio $x'(t) = x(t) + \mu(t)$ that can be used to launch the practical attack over the air.  

Such noise model above is quite device-dependent. Since different speakers and receivers may introduce different noises/distortion when playing or receiving specific audio, $x'(t)$ may only work with the devices that we use to capture the noise. To enhance the robustness of $x'(t)$, we introduce \textbf{\textit{random noise}}, which is shown in Eq~\eqref{eq:modify_2}. Here, the function \textbf{\textit{rand()}} returns an vector of random numbers in the interval (-N,N), which is saved as a ``WAV" format file to represent $n(t)$. Our evaluation results show that this approach can make the adversarial audio $x'(t)$ robust enough for different speakers and receivers.

\begin{equation} \label{eq:modify_2} n(t)=rand(t), \left|n(t)\right|<=N.
\end{equation}

\section{Evaluation}
\label{sec:Evaluation}
In this section, we present the experimental results of CommanderSong. We evaluated both the WTA and WAA attacks against machine recognition. To evaluate the human comprehension, we conducted a survey examining the effects of ``hiding'' the desired command in the song. Then, we tested the transferability of the adversarial sample on other ASR platforms, and checked whether CommanderSong can spread through Internet and radio. Finally, we measured the efficiency in terms of the time to generate the CommanderSong. Demos of attacks are uploaded on the website (https://sites.google.com/view/commandersong/).

\begin{table*}
\centering
\caption{WTA attack results.}
\label{WavtoAPIresults}
\begin{tabular}{m{5.5cm}<{\centering}|m{2.5cm}<{\centering}|m{1.5cm}<{\centering}|m{3.8cm}<{\centering}}
\Xhline{3\arrayrulewidth}
\textbf{Command} & \textbf{Success rate ($\%$)} & \textbf{SNR ($dB$)}& \textbf{Efficiency (frames/hours)}\\ \Xhline{3\arrayrulewidth}
\text{Okay google restart phone now.}& \text{100} & \text{18.6}&  \text{229/1.3}\\ \hline
\text{Okay google flashlight on.} & \text{100} & \text{14.7}& \text{219/1.3}\\\hline
\text{Okay google read mail.} & \text{100} & \text{15.5}& \text{217/1.5}\\ \hline
\text{Okay google clear notification.} & \text{100} &  \text{14}& \text{260/1.2}\\ \hline
\text{Okay google good night.} & \text{100} & \text{15.6}& \text{193/1.3}\\ 
\hline
\text{Okay google airplane mode on.} & \text{100} &  \text{16.9}& \text{219/1.1}\\ \hline
\text{Okay google turn on wireless hot spot.} & \text{100} & \text{14.7}& \text{280/1.6}\\ \hline
\text{Okay google read last sms from boss.} & \text{100}  &  \text{15.1}& \text{323/1.4}\\ \hline
\text{Echo open the front door.} & \text{100} &  \text{17.2}& \text{193/1.0}\\ \hline
\text{Echo turn off the light.} & \text{100} & \text{17.3}& \text{347/1.5}\\ \hline
\multirow{2} {*}\text{Okay google call one one zero one one nine one two zero.} & \text{100} &  \text{14.8}& \text{387/1.7}\\ 
\hline
\multirow{2} {*}\text{Echo ask capital one to make a credit card payment.} & \text{100} &  \text{15.8}& \text{379/1.9}\\ 
\Xhline{3\arrayrulewidth}
\end{tabular}
\end{table*}

\subsection{Experiment Setup}
The pure voice audio of the desired commands can be generated by any Text-To-Speech (TTS) engine (e.g., Google text-to-speech~\cite{GoogleTTS}, etc.) or recording human voice, as long as it can be correctly recognized by Kaldi platform. We also randomly downloaded 26 songs from the Internet. To understand the impact of using different types of songs as the carrier, we intended to choose songs from different categories, i.e., popular, rock, rap, and soft music. Regarding the commands to inject, we chose 12 commonly used ones such as ``turn on GPS'', ``ask Capital One to make a credit card payment'', etc., as shown in Table \ref{WavtoAPIresults}. Regarding the computing environment, one GPU server (1075MHz GPU with 12GB memory, and 512GB hard drive) was used.

\subsection{Effectiveness}
\vspace {3pt}
\noindent\textbf{WTA Attack.} 
In this WTA attack, we directly feed the generated adversarial songs to Kaldi using its exposed APIs, which accept raw audio file as input. Particularly, we injected each command into each of the downloaded 26 songs using the approach proposed in Section \ref{WTA}. Totally we got more than 200 adversarial songs in the ``WAV'' format and sent them to Kaldi directly for recognition. If Kaldi successfully identified the command injected inside, we denote the attack as successful. 

Table~\ref{WavtoAPIresults} shows the WTA attack results. Each command can be recognized by Kaldi correctly. The success rate $100\%$ means Kaldi can decode every word in the desired command correctly. The success rate is calculated as the ratio of the number of words successfully decoded and the number of words in the desired command. Note in the case that the decoded word is only one character different than that in the desired command, we consider the word is not correctly recognized. 

For each adversarial song, we further calculated the average signal-noise ratio (SNR) against the original song as shown in Table \ref{WavtoAPIresults}. SNR is a parameter widely used to quantify the level of a signal power to noise, so we use it here to measure the distortion of the adversarial sample over the original song. We then use the following equation $SNR(dB)=10log_{10}({P_{x(t)}}/{P_{\delta(t)})}$ to obtain SNR, where the original song $x(t)$ is the signal while the perturbation $\delta(t)$ is the noise. Larger SNR value indicates a smaller perturbation. Based on the results in Table \ref{WavtoAPIresults}, the SNR ranges from 14$\thicksim$18.6 $dB$, indicating that the perturbation in the original song is less than 4\%. Therefore, the perturbation should be too slight to be noticed.

\begin{table*}
\centering
\caption{WAA attack results.}
\label{WavairAPIresults}
\begin{tabular}{m{3.2cm}<{\centering}|m{3.1cm}<{\centering}|m{2.5cm}<{\centering}|m{1.5cm}<{\centering}|m{3.8cm}<{\centering}}
\Xhline{3\arrayrulewidth}
\textbf{Command} & \textbf{Speaker}& \textbf{Success rate ($\%$)} & \textbf{SNR ($dB$)} & \textbf{Efficiency (frames/hours)}\\ \Xhline{3\arrayrulewidth}
\text{Echo ask capital one } & {JBL speaker} & \text{90} &\text{1.7}\\ \cline{2-4}
\text{to make a credit card}&\text{ASUS Laptop} & \text{82} &\text{1.7}& \multirow{1} {*}{379/2.0}\\ \cline{2-4}
\text{card payment.}  & {SENMATE Broadcast} & \text{72} &\text{1.7}&\\ \hline
\text{Okay google call one} & {JBL speaker} & \text{96} &\text{1.3}& \\ \cline{2-4}
\text{ one zero one one nine}& \text{ASUS Laptop} & \text{60} &\text{1.3}& \multirow{1} {*}{400/1.8}\\ \cline{2-4}
\text{one two zero.}  & \text{SENMATE Broadcast} & \text{
70
} &\text{1.3}\\ 
\Xhline{3\arrayrulewidth}
\end{tabular}
\end{table*}

\vspace {3pt}
\noindent\textbf{WAA Attack.} 
To practically attack Kaldi over the air, the ideal case is to find a commercial IVC device implemented based on Kaldi and play our adversarial samples against the device. However, we are not aware of any such IVC device, so we simulate a pseudo IVC device based on Kaldi. In particular, the adversarial samples are played by speakers over the air. We use the recording functionality of iPhone 6S to record the audio, which is sent to Kaldi API to decode. Overall, such a pseudo IVC device is built using the microphone in iPhone 6S as the audio recorder, and Kaldi system to decode the audio.

We conducted the practical WAA attack in a meeting room (16 meter long, 8 meter wide, and 4 meter tall). The songs were played using three different speakers including a JBL clip2 portable speaker, an ASUS laptop and a SENMATE broadcast equipment~\cite{SENMATE}, to examine the effectiveness of the injected random noise. All of the speakers are easy to obtain and carry. The distance between the speaker and the pseudo IVC device (i.e., the microphone of the iPhone 6S) was set at 1.5 meters. We chose two commands as in Table~\ref{WavairAPIresults}, and generated adversarial samples. Then we played them over the air using those three different speakers and used the iPhone 6S to record the audios, which were sent to Kaldi to decode. Table~\ref{WavairAPIresults} shows the WAA attack results. For both of the two commands, JBL speaker overwhelms the other two with the success rate up to $96\%$, which might indicate its sound quality is better than the other two. All the SNRs are below 2 $dB$, which indicates slightly bigger perturbation to the original songs due to the random noise from the signal's point of view. Below we will evaluate if such ``bigger'' perturbation is human-noticeable by conducting a survey.

\vspace {3pt}
\noindent\textbf{Human comprehension from the survey.}
To evaluate the effectiveness of hiding the desired command in the song, we conducted a survey on Amazon Mechanical Turk (MTurk)~\cite{AmazonTurk}, an online marketplace for crowdsourcing intelligence. We recruited 204 individuals to participate in our survey\footnote{The survey will not cause any potential risks to the participants (physical, psychological, social, legal, etc.). The questions in our survey do not involve any confidential information about the participants. We obtained the IRB Exempt certificates from our institutes.}. Each participant was asked to listen to 26 adversarial samples, each lasting for about 20 seconds (only about four or five seconds in the middle is crafted to contain the desired command.). A series of questions regarding each audio need to be answered, e.g., (1) whether they have heard the original song before; (2) whether they heard anything abnormal than a regular song (The four options are \textit{no}, \textit{not sure}, \textit{noisy}, and \textit{words different than lyrics}); (3) if choosing \textit{noisy} option in (2), where they believe the noise comes from, while if choosing \textit{words different than lyrics} option in (2), they are asked to write down those words, and how many times they listened to the song before they can recognize the words. 
\begin{table}[!htbp]
\small
\centering
\caption{Human comprehension of the WTA samples.}
\label{Human Comprehension WTA}
\begin{tabular}{m{1.85cm}<{\centering}|m{1.1cm}<{\centering}|m{1.3cm}<{\centering}|m{2.0cm}<{\centering}}
\Xhline{3\arrayrulewidth}
\textbf{Music \quad Classification} & \textbf{Listened ($\%$)} & \textbf{Abnormal ($\%$)} & \textbf{Recognize Command ($\%$) }\\ \Xhline{3\arrayrulewidth}
\text{Soft Music}   & \text{13} &\text{15}&\text{0}\\ 
\hline
\text{Rock}   & \text{33} &\text{28}&\text{0}\\ \hline
\text{Popular}   & \text{32} &\text{26}&\text{0}\\ \hline
\text{Rap}  &  \text{41} &\text{23}&\text{0}\\ 
\Xhline{3\arrayrulewidth}
\end{tabular}
\end{table}

The entire survey lasts for about five to six minutes. Each participant is compensated \$0.3 for successfully completing the study, provided they pass the attention check question to motivate the participants concentrate on the study. Based on our study, 63.7\% of the participants are in the age of 20$\thicksim$40 and 33.3\% are 40$\thicksim$60 years old, and 70.6\% of them use IVC devices (e.g., Amazon Echo, Google home, Smartphone, etc.) everyday. 
\begin{table}[!bp]
\footnotesize
\centering
\caption{Human comprehension of the WAA samples.}
\label{Human Comprehension WAA}
\begin{tabular}{m{1.6cm}<{\centering}|m{1cm}<{\centering}|m{1.2cm}<{\centering}|m{1.0cm}<{\centering}|m{1.0cm}<{\centering}}
\Xhline{3\arrayrulewidth}
\textbf{Song Name}& \textbf{Listened ($\%$)} & \textbf{Abnormal ($\%$)} & \textbf{Noise-speaker ($\%$)}& \textbf{Noise-song ($\%$) }\\ \Xhline{3\arrayrulewidth}
{Did You Need It} & \text{15} &\text{67}&\text{42}&\text{1}\\ \cline{1-5}
{Outlaw of Love} & \text{11} &\text{63}&\text{36}& \text{2}\\ \cline{1-5}
{The Saltwater Room
} & \text{27} &\text{67}& \text{39}&\text{3}\\ \cline{1-5}
\text{Sleepwalker} & \text{13} &\text{67}&\text{41}& \text{0}\\ \cline{1-5}
\text{Underneath} & \text{13} &\text{68} &\text{45}& \text{3}\\ \cline{1-5}
\text{Feeling Good} & \text{38} &\text{59}&\text{36}& \text{4}\\ \Xhline{2\arrayrulewidth}
\textit{Average} & \text{19.5} &\text{65.2}&\text{40}&\text{2.2}\\   
\Xhline{3\arrayrulewidth}
\end{tabular}
\end{table}

Table~\ref{Human Comprehension WTA} shows the results of the human comprehension of our WTA samples. We show the average results for songs belonging to the same category. The detailed results for each individual song can be referred to in Table~\ref{Human Comprehension} in Appendix. Generally, the songs in soft music category are the best candidates for the carrier of the desired command, with as low as 15\% of participants noticed the abnormality. None of the participants could recognize any word of the desired command injected in the adversarial samples of any category. Table~\ref{Human Comprehension WAA} demonstrates the results of the human comprehension of our WAA samples. On average, 40\% of the participants believed the noise was generated by the speaker or like radio, while only 2.2\% of them thought the noise from the samples themselves. In addition, less than 1\% believed that there were other words except the original lyrics. However, none of them successfully identified any word even repeating the songs several times. 

\subsection{Towards the Transferability}
Finally, we assess whether the proposed CommanderSong can be transfered to other ASR platforms. 

\vspace {3pt}
\noindent\textbf{Transfer from Kaldi to iFLYTEK.} We choose iFLYTEK ASR system as the target of our transfer, due to its popularity. As one of the top five ASR systems in the world, it possesses 70\% of the market in China. Some applications supported by iFLYTEK and their downloads on Google Play as well as the number of worldwide users are listed in Table~\ref{iFLYTEK} in Appendix. In particular, \textit{iFLYTEK Input} is a popular mobile voice input method, which supports mandarin, English and personalized input~\cite{iFLYTEKInput}. \textit{iFLYREC} is an online service offered by iFLYTEK to convert audio to text~\cite{iFLYREC}. We use them to test the transferability of our WAA attack samples, and the success rates of different commands are shown in Table~\ref{Kaldi_iFLYTEK}. Note that WAA audio samples are directly fed to \textit{iFLYREC} to decode. Meanwhile, they are played using Bose Companion 2 speaker towards \textit{iFLYTEK Input} running on smartphone LG V20, or using JBL speaker towards \textit{iFLYTEK Input} running on smartphone Huawei honor 8/MI note3/iPhone 6S. Those adversarial samples containing commands like \textit{open the door} or \textit{good night} can achieve great transferability on both platforms. However, the command \textit{airplane mode on} only gets 66\% success rate on \textit{iFLYREC}, and 0 on \textit{iFLYTEK Input}.

\vspace {3pt}
\noindent\textbf{Transferability from Kaldi to DeepSpeech.}
We also try to transfer CommanderSong from Kaldi to DeepSpeech, which is an open source end-to-end ASR system. We directly fed several adversarial WTA and WAA attack samples to DeepSpeech, but none of them can be decoded correctly. As Carlini et al. have successfully modified any audio into a command recognizable by DeepSpeech~\cite{AttackDeepSpeech}, we intend to leverage their open source algorithm to examine if it is possible to generate one adversarial sample against both two platforms. In this experiment, we started by 10 adversarial samples generated by CommanderSong, either WTA or WAA attack, integrating commands like \textit{Okay google call one one zero one one nine one two zero}, \textit{Echo open the front door}, and \textit{Echo turn off the light}. We applied their algorithm to modify the samples until DeepSpeech can decode the target commands correctly. Then we tested such
newly generated samples against Kaldi as WTA attack,
and Kaldi can still successfully recognize them. We did
not perform WAA attack since their algorithm targeting
DeepSpeech cannot achieve attacks over the air.

The preliminary evaluations on transferability give us
the opportunities to understand CommanderSongs and for
designing systematic approach to transfer in the future.
\subsection{Automated Spreading}
Since our WAA attack samples can be used to launch the practical adversarial attack against ASR systems, we want to explore the potential channels that can be leveraged to impact a large amount of victims automatically.
\begin{table}
\centering
\caption{Transferability from Kaldi to iFLYTEK.}
\label{Kaldi_iFLYTEK}
\begin{tabular}{m{2.6cm}<{\centering}|m{2.0cm}<{\centering}|m{1.5cm}<{\centering}}
\Xhline{3\arrayrulewidth}
\textbf{Command} & \textbf{iFLYREC ($\%$)} & \textbf{iFLYTEK Input ($\%$)}\\ \Xhline{3\arrayrulewidth}
\text{Airplane mode on.}& \text{66} & \text{0}\\ \hline
\text{Open the door.} & \text{100} &  \text{100}\\ \hline
\text{Good night.} & \text{100} & \text{100}\\ \hline
\Xhline{3\arrayrulewidth}
\end{tabular}
\end{table}

\vspace {3pt}
\noindent\textbf{Online sharing.} We consider the online sharing platforms like YouTube to spread CommanderSong. We picked up one five-second adversarial sample embedded with the command ``\textit{open the door}'' and applied Windows Movie Maker software to make a video, since YouTube only supports video uploading. The sample was repeated four times to make the full video around 20 seconds. We then connected our desktop audio output to Bose Companion 2 speaker and installed \textit{iFLYTEK Input} on LG V20 smartphone. In this experiment, the distance between the speaker and the phone can be up to 0.5 meter, and \textit{iFLYTEK Input} can still decode the command successfully. 

\vspace {3pt}
\noindent\textbf{Radio broadcasting.} In this experiment, we used HackRF One~\cite{HackRF}, a hardware that supports Software Defined Radio (SDR) to broadcast our CommanderSong at the frequency of FM 103.4 MHz, simulating a radio station. We setup a radio at the corresponding frequency, so it can receive and play the CommanderSong. We ran the \textit{WeChat}\footnote{\textit{WeChat} is the most popular instant messaging application in China, with approximately 963,000,000 users all over the world by June 2017~\cite{wechat}.} application and enabled the \textit{iFLYTEK Input} on different smartphones including iPhone 6S, Huawei Honor 8 and XiaoMi MI Note3. \textit{iFLYTEK Input} can always successfully recognize the command ``\textit{open the door}'' from the audio played by the radio and display it on the screen.

\subsection{Efficiency}
We also evaluate the cost of generating CommanderSong in the aspect of the required time. For each command, we record the time to inject it into different songs and compute the average. Since the time required to craft also depends on the length of the desired command, we define the efficiency as the ratio of the number of frames of the desired command and the required time. Table~\ref{WavtoAPIresults} and Table~\ref{WavairAPIresults} show the efficiency of generating WTA and WAA samples for different commands. Most of those adversarial samples can be generated in less than two hours, and some simple commands like ``\textit{Echo open the front door}'' can be done within half an hour. However, we do notice that some special words (such as \textit{GPS} and \textit{airplane}) in the command make the generation time longer. Probably those words are not commonly used in the training process of the ``ASpIRE model'' of Kaldi, so generating enough phonemes to represent the words is time-consuming. Furthermore, we find that, for some songs in the rock category such as ``\textit{Bang bang}'' and ``\textit{Roaked}'', it usually takes longer to generate the adversarial samples for the same command compared with the songs in other categories, probably due to the unstable rhythm of them.

\begin{figure}[t]
\centering
 \includegraphics[width=0.48\textwidth]{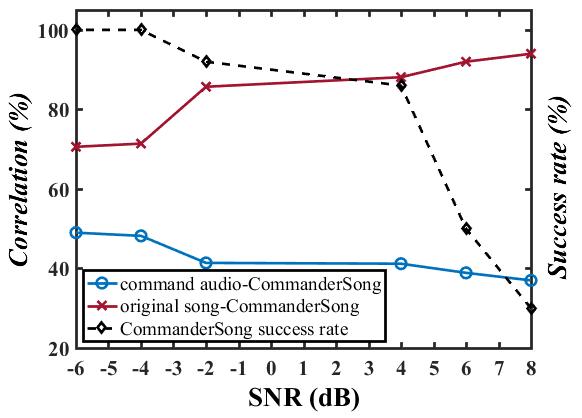}
\caption{SNR impacts on correlation of the audios and the success rate of adversarial audios.  
\label{fig_exp_mat}}
\end{figure}

\section{Understanding the Attacks}
\label{sec:understanding}

We try to deeply understand the attacks, which could potentially help to derive defense approaches. We raise some questions and perform further analysis on the attacks. 

\vspace{3pt}\noindent\textbf{In what ways does the song help the attack?} We use songs as the carrier of commands to attack ASR systems. Obviously, one benefit of using a song is to prevent listeners from being aware of the attack. Also CommanderSong can be easily spread through Youtube, radio, TV, etc. Does the song itself help generate the adversarial audio samples? To answer this question, we use a piece of silent audio as the ``carrier" to generate CommanderSong $A_{cs}$ (WAA attack), and test the effectiveness of it. The results show that $A_{cs}$ can work, which is aligned with our findings -- a random song can serve as the ``carrier" because a piece of silent audio can be viewed as a special song. 

However, after listening to $A_{cs}$, we find that $A_{cs}$ sounds quite similar to the injected command, which means any user can easily notice it, so $A_{cs}$ is not the adversarial samples we desire. Note that, in our human subject study, none of the participants recognized any command from the generated CommanderSongs. We assume that \textit{some phonemes or even smaller units in the original song work together with the injected small perturbations to form the target command}. To verify this assumption, we prepare a song $A_s$ and use it to generate the CommanderSong $A_{cs}$. Then we calculate the difference $\Delta(A_{s}, A_{cs})$ between them, and try to attack ASR systems using $\Delta(A_{s}, A_{cs})$. However, after several times of testing, we find that $\Delta(A_{s}, A_{cs})$ does not work, which indicates the pure perturbations we injected cannot be recognized as the target commands.

Recall that in Table~\ref{Human Comprehension WAA}, the songs in the soft music category are proven to be the best carrier, with lowest abnormality identified by participants. Based on the findings above, it appears that such songs can better aligned with the phonemes or smaller ``units" in the target command to help the attack. This is also the reason why $\Delta(A_{s}, A_{cs})$ cannot directly attack successfully: the ``units" in the song combined with $\Delta(A_{s}, A_{cs})$ together construct the phonemes of the target command.

\begin{figure}[t]
\centering
\includegraphics[width=0.48\textwidth]{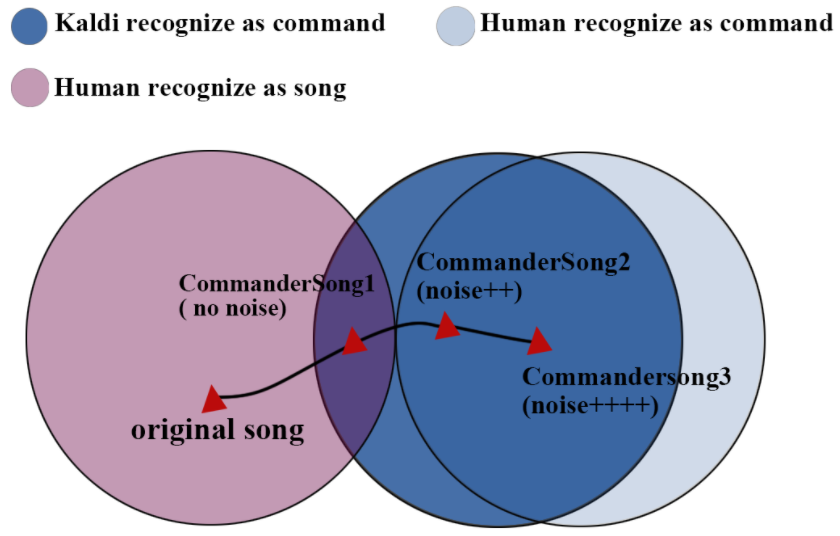}
\caption{Explaination of Kaldi and human recognition of the audios.  
\label{Explain}}
\end{figure}

\vspace{3pt}\noindent\textbf{What is the impact of noise in generating adversarial samples?} As mentioned early, we build a generic random noise model to perform the WAA attack over the air. In order to understand the impact of the noise in generating adversarial samples, we crafted CommanderSong using noises with different amplitude values. Then we observed the differences between the CommanderSong and the original song, the differences between the CommanderSong and the pure command audio, and the success rates of the CommanderSong to attack. To characterize the difference, we leverage Spearman's rank correlation coefficient~\cite{pirie1988spearman} (\textit{Spearman's rho} for short) to represent the similarity between two pieces of audio. Spearman's rho is widely used to represent the correlation between two variables, and can be calculated as follows: $r(X,Y)=Cov(X,Y)/\sqrt{Var[X]Var[Y]}$, where $X$ and $Y$ are the MFCC features of the two pieces of audio. $Cov(X,Y)$ represents the covariance of X and Y. $Var[X]$ and $Var[Y]$ are the variances of X and Y respectively. 

The results are shown in Figure~\ref{fig_exp_mat}. The x-axis in the figure shows the $SNR$ (in $dB$) of the noise, and the y-axis gives the correlation. From the figure, we find that the correlation between the CommanderSong and the original song (red line) decreases with $SNR$. It means that the CommanderSong sounds less like the original song when the amplitude value of the noise becomes larger. This is mainly because the original song has to be modified more to find a CommanderSong robust enough against the introduced noise. On the contrary, the CommanderSong becomes more similar with the target command audio when the amplitude values of the noise increases (i.e., decrease of $SNR$ in the figure, blue line), which means that the CommanderSong sounds more like the target command. The success rate (black dotted line) also increases with the decrease of SNR. We also note that, when $SNR=4$ $dB$, the success rate could be as high as 88\%. Also the correlation between CommanderSong and the original song is 90\%, which indicates a high similarity.

\begin{figure}[t]
\centering
\includegraphics[width=0.48\textwidth]{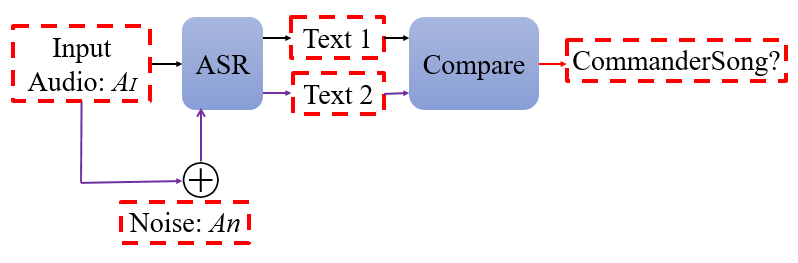}
\caption{Audio turbulence defense.
\label{defense1}}
\end{figure}

Figure~\ref{Explain} shows the results from another perspective. Suppose the dark blue circle is the set of audios that can be recognized as commands by ASR systems, while the light blue circle and the red one represent the sets of audio recognized as commands and songs by human respectively. At first, the original song is in the red circle, which means that neither ASR systems nor human being recognize any command inside. WTA attack slightly modifies the song so that the open source system Kaldi can recognize the command while human cannot. After noises are introduced to generate CommanderSong for WAA attacks, CommanderSong will fall into the light blue area step by step, and in the end be recognized by human. Therefore, attackers can choose the amplitude values of noise to balance between robustness to noise and identifiability by human users.

\section{Defense}
\label{sec:Defense}
We propose two approaches to defend against CommanderSong: Audio turbulence and Audio squeezing. The first defense is effective against WTA, but not WAA; while the second defense works against both attacks.

\vspace{3pt}\noindent\textbf{Audio turbulence.} From the evaluation, we observe that noise (e.g., from speaker or background) decreases the success rate of CommanderSong while impacts little on the recognition of audio command. So our basic idea is to add noise (referred to as \textit{turbulence} noise $A_n$) to the input audio $A_I$ before it is received by the ASR system, and check whether the resultant audio $A_I \textcircled{+} A_n$ can be interpreted as other words. Particularly, as shown in Figure~\ref{defense1}, $A_I$ is decoded as \texttt{text}$_1$ by the ASR system. Then we add $A_n$ to $A_I$ and let the ASR system extract the text \texttt{text}$_2$ from $A_I \textcircled{+} A_n$. If \texttt{text}$_1 \neq$\texttt{text}$_2$, we say that the CommanderSong is detected.
 
\begin{figure}[t]
\centering
\includegraphics[width=0.48\textwidth]{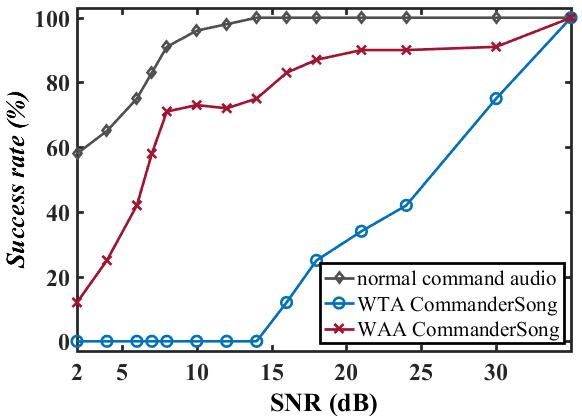}
\caption{The results of audio turbulence defense.  
\label{Audioturbulence_defense}}
\end{figure}

We did experiments using this approach to test the effectiveness of such defense. The target command ``open the door'' was used to generate a CommanderSong. Figure~\ref{Audioturbulence_defense} shows the result. The x-axis shows the $SNR$ ($A_I$ to $A_n$), and the y-axis shows the success rate. We found that the success rate of WTA dramatically drops when $SNR$ decreases. When $SNR=15$ $dB$, WTA almost always fails and $A_I$ can still be successfully recognized, which means this approach works for WTA. However, the success rate of WAA is still very high. This is mainly because CommanderSongs for WAA is generated using random noises, which is robust against turbulence noise.

\begin{figure}[t]
\centering
\includegraphics[width=0.48\textwidth]{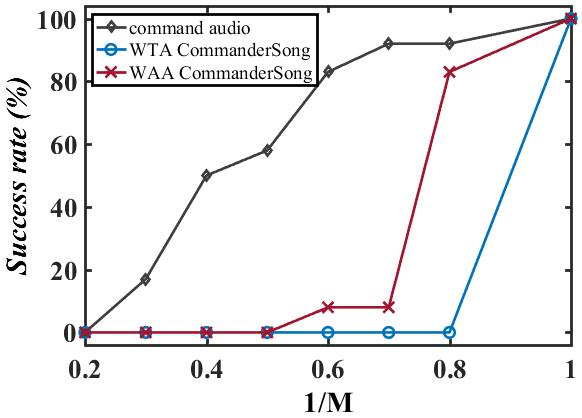}
\caption{Audio squeezing defense result.  
\label{Audiosqueezing_defense}}
\end{figure}

\vspace{3pt}\noindent\textbf{Audio squeezing.} The second defense is to reduce the sampling rate of the input audio $A_I$ (just like squeezing the audio). Instead of adding $A_n$ in the defense of audio turbulence, we downsample $A_I$ (referred to as $D(A_I)$). Still, ASR systems decode $A_I$ and $D(A_I)$, and get \texttt{text}$_1$ and \texttt{text}$_2$ respectively. If \texttt{text}$_1 \neq$\texttt{text}$_2$, the CommanderSong is detected. Similar to the previous experiment, we evaluate the effectiveness of this approach. The results are shown in Figure~\ref{Audiosqueezing_defense}. The x-axis shows the ratio ($1/M$) of downsampling (M is the downsampling factor or decimation factor, which means that the original sampling rate is M times of the downsampled rate). When $1/M=0.7$ (if the sample rate is 8000 samples/second, the downsampled rate is 5600 samples/second), the success rates of WTA and WAA are 0\% and 8\% respectively. $A_I$ can still be successful recognized at the rate of 91\%. This means that Audio squeezing is effective to defend against both WTA and WAA.

\section{Related Work}
\label{sec:relate}
\vspace {3pt}
\noindent\textbf{Attack on ASR system.} Prior to our work, many researchers have devoted to security issues about speech controllable systems~\cite{GhostTalk, kasmi2015iemi,diao2014your,mukhopadhyay2015all,CocaineNoodles,carlini2016hidden,zhang2017dolphinattack,AttackDeepSpeech}. Denis et al. found the vulnerability of analog sensor and injected bogus voice signal to attack the microphone~\cite{GhostTalk}. Kasmi et al. stated that, by leveraging intentional electromagnetic interference on headset cables, voice command could be injected and carried by FM signals which is further received and interpreted by smart phones~\cite{kasmi2015iemi}. Diao et al. demonstrated that, through permission bypassing attack in Android smart phones, voice commands could be played using apps with zero permissions~\cite{diao2014your}. Mukhopadhyay et al. considered voice impersonation attacks to contaminate a voice-based user authentication system~\cite{mukhopadhyay2015all}. They reconstructed the victim’s voice model from the victim’s voice data, and launched attacks that can bypass voice authentication systems. Different from these attacks, we are attacking the machine learning models of ASR systems.

Hidden voice command~\cite{carlini2016hidden} launched both black box (i.e., inverse MFCC) and white box (i.e., gradient decent) attacks against ASR systems with GMM-based acoustic models. Different from this work, our target is a DNN-based ASR system. Recently, the authors posted the achievement that can construct targeted audio adversarial examples on DeepSpeech, an end-to-end open source ASR platform~\cite{AttackDeepSpeech}. To perform the attack, the adversary needs to directly upload the adversarial WAV file to the speech recognition system. Our attacks on Kaldi are concurrent to their work, and our attack approaches are independent to theirs. Moreover, our attacks succeed under a more practical setting that let the adversarial audio be played over the air. The recent work DolphinAttack~\cite{zhang2017dolphinattack} proposed a completely inaudible voice attack by modulating commands on ultrasound carriers and leveraging microphone vulnerabilities to attack. As noted by the authors, such attack can be eliminated by filtering out ultrasound carrier (e.g., iPhone 6 Plus). Differently, our attack uses songs instead of ultrasound as the carriers, making the attack harder to defend.

\vspace {3pt}
\noindent\textbf{Adversarial research on machine learning.} Besides attacking speech recognition systems, there has been substantial work on adversarial machine learning examples towards physical world. Kurakin et al. ~\cite{kurakin2016adversarial} proved it is doable that Inception v3 image classification neural network could be compromised by adversarial images. Brown et al. ~\cite{brown2017adversarial} showed by adding an universal patch to an image they could fool the image classifiers successfully. Evtimov et al. ~\cite{evtimov2017robust} proposed a general algorithm which can produce robust adversarial perturbations into images to overcome physical condition in real world. They successfully fooled road sign classifiers to mis-classify real Stop Sign. Different from them, our study targets speech recognition system.

\vspace {3pt}
\noindent\textbf{Defense of Adversarial on machine learning.} Defending against adversarial attacks is known to be a challenging problem. Existing defenses include adversarial training and defensive distillation. Adversarial training~\cite{madry2017towards} adds the adversarial examples into the model's training set to increase its robustness against these examples. Defensive distillation~\cite{hinton2015distilling} trains the model with probabilities of different class labels supported by an early model trained on the same task. Both defenses perform a kind of gradient masking~\cite{papernot2017practical} which increases the difficulties for the adversary to compute the gradient direction. In~\cite{DawnSong}, Dawn Song attempted to combine multiple defenses including feature squeezing and the specialist to construct a larger strong defense. They stated that defenses should be evaluated by strong attacks and adaptive adversarial examples. Most of these defenses are effective for white box attacks but not for black box ones. Binary classification is another simple and effective defense for white box attacks without any modifications of the underlying systems. A binary classifier is built to separate adversarial examples apart from the clean data. Similar as adversarial training and defensive distillation, this defense suffers from generalization limitation. In this paper, we propose two novel defenses against CommanderSong attack.

\section{Conclusion}
\label{sec:Conclusion}
In this paper, we perform practical adversarial attacks on ASR systems by injecting ``voice" commands into songs (CommanderSong). To the best of our knowledge, this is the first systematical approach to generate such practical attacks against DNN-based ASR system. Such CommanderSong could let ASR systems execute the command while being played over the air without notice by users. Our evaluation shows that CommanderSong can be transferred to iFLYTEK, impacting popular apps such as WeChat, Sina Weibo, and JD with billions of users. We also demonstrated that CommanderSong can be spread through YouTube and radio. Two approaches (audio turbulence and audio squeezing) are proposed to defend against CommanderSong.

\section*{Acknowledgments}
IIE authors are supported in part by National Key R\&D Program of China (No.2016QY04W0805), NSFC U1536106, 61728209, National Top-notch Youth Talents Program of China, Youth Innovation Promotion Association CAS and Beijing Nova Program. 
Indiana University author is supported in part by the NSF 1408874, 1527141, 1618493 and ARO W911NF1610127. Illinois University authors are supported in part by NSF CNS grants 13-30491, 14-08944, and 15-13939.

{\footnotesize\bibliographystyle{plain}
\bibliography{bibliography}
\clearpage
\onecolumn
\begin{appendices}  
\section*{Appendix }
\begin{table*}[!htbp]
\centering
\caption{The detailed results of individual song in human comprehension survey for WTA samples. When we were checking the survey results from MTurk, we found the average familiarity of MTurk workers towards our songs is not as good as we expected. So streaming counts from Spotify are also listed in the table, as we want to show the popularity of our sample songs. The song \textit{Selling Brick in Street} is not in Spotify database so we can not provide the count for it.}
\label{Human Comprehension}
\begin{tabular}{m{1.8cm}<{\centering}|m{4.95cm}<{\centering}|m{2.58cm}<{\centering}|m{1.1cm}<{\centering}|m{1.4cm}<{\centering}|m{2.2cm}<{\centering}}
\Xhline{3\arrayrulewidth}
\textbf{Music Classification} & \textbf{Song Name}& \textbf{Spotify Streaming Count}&\textbf{Listened ($\%$)} & \textbf{Abnormal ($\%$)} & \textbf{Recognize Command ($\%$) }\\ \Xhline{3\arrayrulewidth}
\text{} & {Heart and Soul} & {13,749,471}&\text{15$\%$} &\text{8$\%$}&\text{0}\\ \cline{2-5}
\text{}&\text{Castle in the Sky} & {2,332,348}&\text{9$\%$} &\text{6$\%$}& \text{0}\\ \cline{2-5}
\text{Soft Music} & {A Comme Amour} & {1,878,899}&\text{14$\%$} &\text{18$\%$}& \text{0}\\ \cline{2-5}
\text{}& \text{Mariage D'amour} & {337,486}&\text{17$\%$} &\text{33$\%$}& \text{0}\\ \cline{2-5}
\text{}  & {Lotus} & {49,443,256}&\text{11$\%$} &\text{12$\%$}&\text{0}\\ \cline{2-5}
\text{}  & \textit{Average} & {13,548,292}&\text{13$\%$} &\text{15$\%$}&\text{0}\\ 
\hline
\text{} & {Bang Bang} &{532,057,658}& \text{52$\%$} &\text{24$\%$}&\text{0} \\ \cline{2-5}
\text{}& \text{Soaked} &{29,734}& \text{13$\%$} &\text{32$\%$}& \text{0}\\ \cline{2-5}
\text{Rock} & {Gold} & {11,614,629}&\text{14$\%$} &\text{41$\%$}&\text{0} \\ \cline{2-5}
\text{}& \text{We are never Getting back together} &{113,806,946}& \text{66$\%$} &\text{38$\%$}& \text{0}\\ \cline{2-5}
\text{}  & \text{When can I See You again} & {26,463,993}&\text{20$\%$} &\text{9$\%$}&\text{0}\\ 
\cline{2-5}
\text{}  & \textit{Average} &{136,794,562}& \text{33$\%$} &\text{28$\%$}&\text{0}\\ 
\hline
\text{}& \text{Love Story} &{109,952,344}& \text{49$\%$} &\text{24$\%$}& \text{0}\\ \cline{2-5}
\text{} & {Hello Seattle} &{9,850,328}& \text{29$\%$} &\text{16$\%$}& \text{0}\\
\cline{2-5}
\text{Popular} & {Good Time} &{125,125,693}& \text{48$\%$} &\text{32$\%$}&\text{0} \\ \cline{2-5}
\text{}& \text{To the Sky} & {4,860,627}&\text{27$\%$} &\text{30$\%$}& \text{0}\\ \cline{2-5}
\text{} & {A Loaded Smile} &{658,814}& \text{8$\%$} &\text{26$\%$}&\text{0} \\ \cline{2-5}
\text{}  & \textit{Average} &{50,089,561}& \text{32$\%$} &\text{26$\%$}&\text{0}\\ \hline
\text{}& \text{Rap God} &{349,754,768}& \text{43$\%$} &\text{32$\%$}& \text{0}\\ \cline{2-5}
\text{} & {Let Me Hold You} &{311,569,726}& \text{31$\%$} &\text{15$\%$}& \text{0}\\
\cline{2-5}
\text{Rap} & {Lose Yourself} & {483,937,007}&\text{75$\%$} &\text{14$\%$}&\text{0} \\ \cline{2-5}
\text{}& \text{Remember the Name} & {193,564,886}&\text{48$\%$} &\text{32$\%$}& \text{0}\\ \cline{2-5}
\text{} & {Selling Brick in Street} &{N/A}& \text{6$\%$} &\text{24$\%$}&\text{0} \\ \cline{2-5}
\text{}  & \textit{Average} & {334,706,597}&\text{41$\%$} &\text{23$\%$}&\text{0}\\ 
\Xhline{3\arrayrulewidth}
\end{tabular}
\end{table*}

\begin{table*}[!htbp]
\centering
\caption{
The detailed information of some sample applications which utilize iFLYTEK as voice input, including number of downloads from Google Play and total user amount. Since Google Services are not accessible in China and information of Apple App Store is not collected, the number of users may not be associated with the number of downloads in Google Play. As shown in the table, each of these applications has over 0.2 billion users in the world.}
\label{iFLYTEK}
\begin{tabular}{m{3cm}<{\centering}|m{3cm}<{\centering}|m{2.8cm}<{\centering}|m{4cm}<{\centering}} 
\Xhline{3\arrayrulewidth}
\textbf{Application} & \textbf{Usage} & \textbf{Downloads from Google Play}& \textbf{Total Users Worldwide (Billion)} \\ \Xhline{3\arrayrulewidth}
\text{Sina Weibo} & \text{Social platform} &\text{11,000,000} &{0.53} \\ \hline
\text{JD} & \text{Online shopping} & \text{1,000,000}&\text{0.27} \\ \hline
\text{CMbrowser}& \text{Searching engine} & \text{50,000,000}&\text{0.64} \\ \hline
\text{Ctrip} & \text{Travel advice website} & \text{1,000,000}&\text{0.30} \\ \hline
\text{Migu Digital} & \text{Voice assistant} & \text{5,000}&\text{0.46} \\ \hline
\text{WeChat} & \text{Chatting, Social} & \text{100,000,000}&\text{0.96} \\ \hline
\text{iFLYTEK Input} & \text{Typing, Voice Input} & \text{500,000}&\text{0.5} \\ 
\Xhline{3\arrayrulewidth}
\end{tabular}
\end{table*}
\end{appendices} 


\end{document}